\documentclass[journal=nalefd,manuscript=letter]{achemso}
\usepackage[version=3]{mhchem} 
\usepackage{caption}
\usepackage{subcaption}
\usepackage{indentfirst}

\author{Yiou Zhang}
\affiliation{Department of Physics, Emory University, Atlanta, Georgia 30322, USA}
\email{yiou.zhang@emory.edu}
\author{Shashi Pandey}
\affiliation{Department of Physics and Astronomy, University of Tennessee, Knoxville, Tennessee 37996, USA}
\author{Sergei Ivanov}
\affiliation{Department of Physics, Emory University, Atlanta, Georgia 30322, USA}
\author{Jian Liu}
\affiliation{Department of Physics and Astronomy, University of Tennessee, Knoxville, Tennessee 37996, USA}
\author{Sergei Urazhdin}
\affiliation{Department of Physics, Emory University, Atlanta, Georgia 30322, USA}

\title{Shot noise in a metal close to Mott transition}


\begin{document}

\begin{abstract}  
  
SrIrO$_3$ is a metallic complex oxide with unusual electronic and magnetic properties believed to originate from electron correlations due to its proximity to Mott metal-insulator transition. However, the nature of its electronic state and the mechanism of metallic conduction remain poorly understood. We demonstrate that shot noise produced by nanoscale SrIrO$_3$ junctions is strongly suppressed, inconsistent with diffusive quasiparticle transport. Analysis of thermal effects and scaling with the junction length reveals that conduction is mediated by collective hopping of electrons almost localized by correlations. Our results provide insight into the non-Fermi liquid state close to Mott transition, and advance shot noise measurements as a powerful technique for the studies of quantum materials.

\end{abstract}

\break

\textbf{Introduction}. Interplay among crystal field effects, electron interactions, and spin-orbit coupling (SOC) leads to rich phenomenology of the quantum states of matter such as topological Mott insulators~\cite{pesin2010mott}, unconventional superconductors~\cite{wang2011twisted}, quantum spin and anomalous Hall effect insulators~\cite{Konig2007,Chang2013}. Complex oxides are archetypal quantum materials whose properties are tunable by varying composition and doping, serving as a test bed for the underlying mechanisms~\cite{Dagotto2005}. Ruddlesden-Popper series of iridium oxides Sr$_{n+1}$Ir$_{n}$O$_{3n+1}$ provide a unique insight into the effects of dimensionality and large SOC on electron correlations. The $5d$ levels of Ir are split by the crystal field and SOC, resulting in a half-filled spin-orbit coupled $J_{eff}=1/2$ valence state whose properties can be approximated by the Mott-Hubbard model of metal-insulator transition (MIT)~\cite{kim2008novel,zeb2012interplay,carter2013theory}. According to this model, the MIT is controlled by the ratio $W/U$ of the effective band width $W$ to the Mott interaction $U$~\cite{zeb2012interplay,moon2008dimensionality,carter2013theory}. In iridates, slabs of $n$ perovskite layers are separated by the ``blocking" SrO layers, allowing effective bandwidth tuning. Indeed, quasi-2D Sr$_2$IrO$_4$ ($n=1$) and Sr$_3$Ir$_2$O$_7$ ($n=2$) are Mott insulators, while SIO ($n=\infty$) is a conducting material which based on angular resolved photoemission spectroscopy (ARPES) and \textit{ab initio} calculations is believed to be a semimetal~\cite{zhang2018review,zeb2012interplay,biswas2014metal,carter2013theory,moon2008dimensionality}. Thus, SIO is a metal (or a semimetal) close to MIT, in the sense that Mott criterion for electron localization is almost satisfied.

Proximity to Mott transition is evidenced by the anomalous transport properties of SIO films. The resistivity $\rho$ of thick films is typically almost independent of temperature $T$. In thin films, it starts to increase with decreasing  $T$~\cite{zhang2018review,biswas2016metal}. Ultrathin films exhibit MIT, consistent with the effects of reduced dimensionality in quasi-2D iridates~\cite{biswas2014metal,groenendijk2017spin,schutz2017dimensionality}. Transport properties of moderately thin SIO films were interpreted in terms of weak localization~\cite{zhang2015tunable,gruenewald2014compressive,biswas2014metal}, and those of ultrathin films $-$ in terms of variable-range-hopping (VRH) and strong (Anderson) localization due to disorder~\cite{biswas2014metal,groenendijk2017spin,zhang2014sensitively}. 
However, the dependence of effective VRH dimensionality on film thickness~\cite{zhang2014sensitively} and inconsistencies between the effects of strain and film thickness on magnetoresistance~\cite{biswas2014metal} indicate significant electron correlations not captured by these models. Indeed, observations of magnetism in thin SIO films and at the interfaces with ferromagnets were interpreted as evidence for significant Mott correlations~\cite{Matsuno2016,Yoo2021,chaurasia2021low}. Meanwhile, electronic Raman spectroscopy shows spectral continuum similar to that in superconducting cuprates, inconsistent with the single-particle Fermi liquid picture~\cite{Sen2020}.

To elucidate the nature of the electronic state in SIO, we performed measurements of shot noise (SN) $-$ white noise produced by biased junctions due to the discrete nature of charge carriers. The current noise in tunnel junctions is~\cite{schottky1918regarding}
\begin{equation}
  S_I = 2qI,
  \label{eqn:shot_current}
\end{equation}
where $q$ is the effective carrier charge and $I$ is the bias current. The Fano factor $F=\frac{S_I}{2eI}$ provides a unique probe for electron correlations, as demonstrated for Cooper pairing in superconductors~\cite{lefloch2003doubled,bastiaans2019imaging} and charge fractionalization in quantum Hall systems~\cite{de1998direct,saminadayar1997observation}. SN is suppressed in mesoscopic conductors. For diffusive single-electron transport in Fermi liquids, Fano factor is reduced to $1/3$ if electron interaction is weak, and $\sqrt{3}/4$ if electrons are thermalized by  interactions~\cite{nagaev1992shot,nagaev1995influence}. The value of $F$ is independent of the conductor length $L$ if the latter is smaller than the electron-phonon scattering length $l_{e-ph}$, and becomes suppressed by electron-phonon interaction in longer junctions~\cite{steinbach1996observation}.

Strong suppression of SN in a non-Fermi liquid ``strange'' metal was attributed to the lack of well-defined quasiparticles~\cite{chen2023shot}. Since the ``strange'' metal state is often observed close to Mott MIT, SIO may be conjectured to also exhibit anomalous SN. Here, we report strong SN suppression in SIO junctions, in agreement with the main result of Ref.~\cite{chen2023shot}. Analysis of thermal effects and scaling with $L$ allows us to eliminate electron-phonon scattering as the SN suppression mechanism, and shows that charge transport is mediated not by quasiparticle diffusion, but by correlated hopping of electrons almost localized by interactions. Our results provide insight into the nature of non-Fermi liquid state close to Mott MIT.

\begin{figure}
	\centering
	\begin{subfigure}[b]{0.45\textwidth}
		\includegraphics[width=\textwidth]{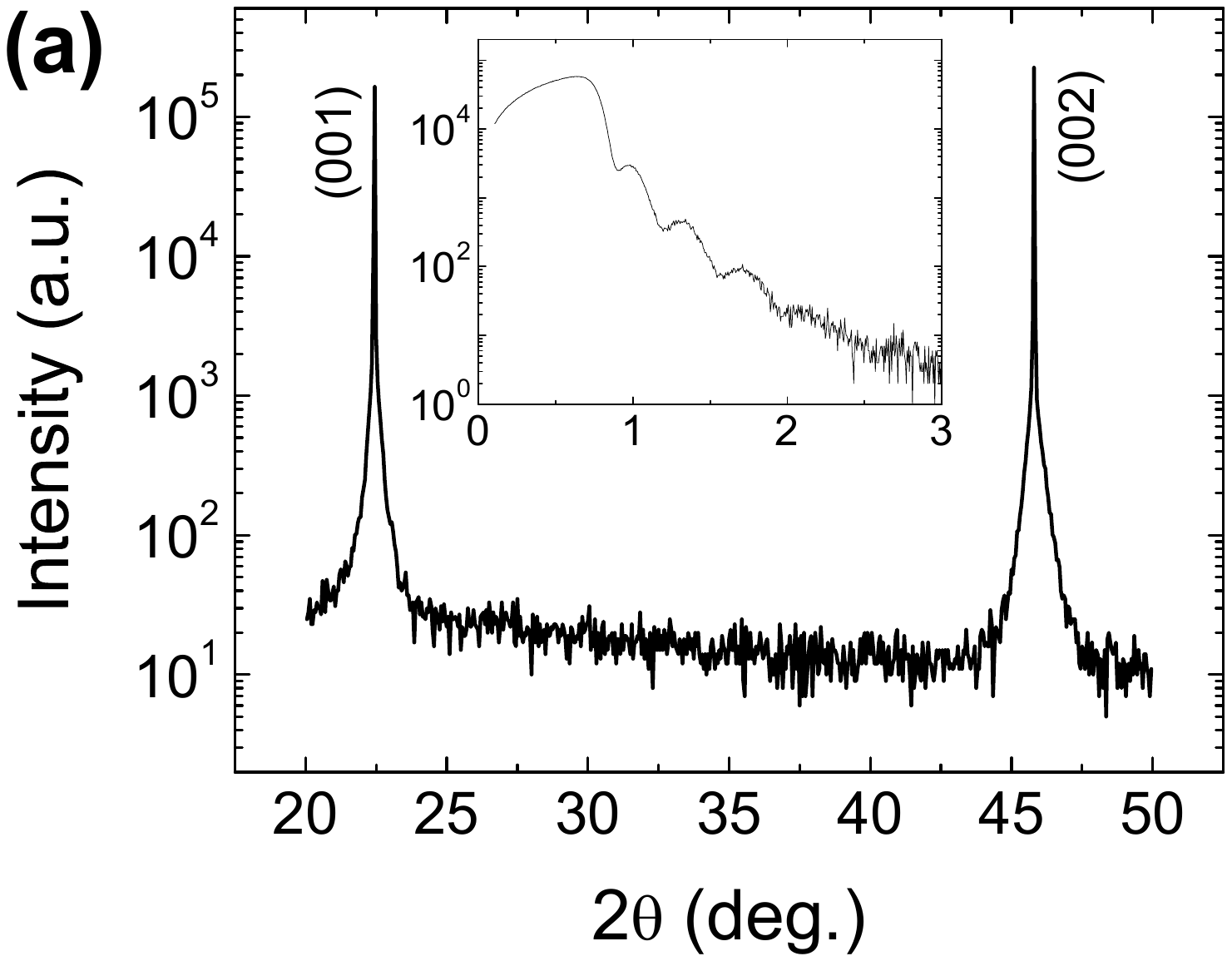}
	\end{subfigure}
	\begin{subfigure}[b]{0.45\textwidth}
		\includegraphics[width=\textwidth]{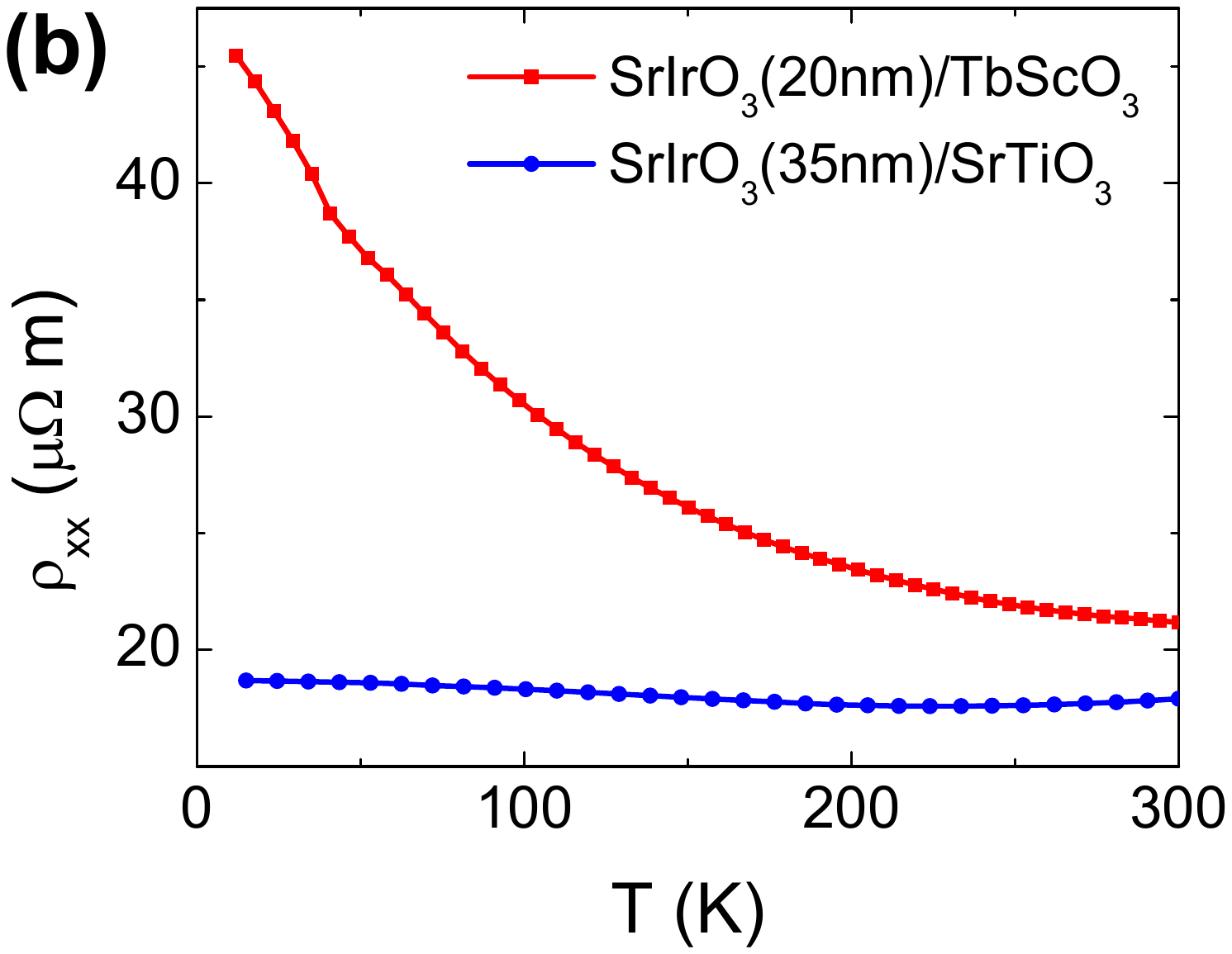}
	\end{subfigure}
	\caption{(a) X-ray diffraction pattern around the TbScO3 (001) and (002) reflections for the SIO(20nm)/TbScO3 film. The film peaks are overlapping with the substrate because of the similar pseudocubic lattice parameters between SIO and TbScO$_3$. The inset shows Kiessig fringes in low-angle X-ray reflectivity, demonstrating low surface/interface roughness. (b) Resistivity vs temperature for unpatterned SrIrO$_3$(20)/TbScO$_3$ and SrIrO$_3$(35)/SrTiO$_3$ films.}
	\label{fgr:fig1}
\end{figure}

\textbf{Results}.  The studied nanostructures were patterned from two films: a SIO(20) film grown on (001)-oriented single-crystalline TbScO$_3$, and a SIO(35) film grown on (001)-oriented SrTiO$_3$. Numbers in parentheses are thicknesses in nanometers. The consistency of results for two different substrates and film thicknesses confirm their generality. The films were deposited from a SrIrO$_3$ target by pulsed laser deposition, with the substrate temperature maintained at $700^\circ$~C, the KrF excimer laser fluence of $2$~J/cm$^2$, and oxygen pressure of $0.1$~mbar. The growth was monitored by in-situ reflection high-energy electron diffraction.  The high quality of epitaxy was confirmed by the sharp peaks in x-ray diffraction measurements, Fig.~\ref{fgr:fig1}(a) (see also Supplmental Materials). Pronounced Kiessig fringes in low-angle reflectivity indicate smooth film interfaces (inset in Fig.~\ref{fgr:fig1}(a)). 

The resistivity $\rho(T)$ of SIO(20) on TbScO$_3$ exhibits a modest non-diverging increase with decreasing $T$, Fig.~\ref{fgr:fig1}(b). Meanwhile, for SIO(35) on SrTiO$_3$, $\rho$ is smaller and almost independent of $T$, which can be attributed to hopping enhancement (larger effective $W$) due to compressive strain and larger film thickness~\cite{gruenewald2014compressive}. 

\begin{figure}
	\centering
	\begin{subfigure}[b]{0.85\textwidth}
		\includegraphics[width=\textwidth]{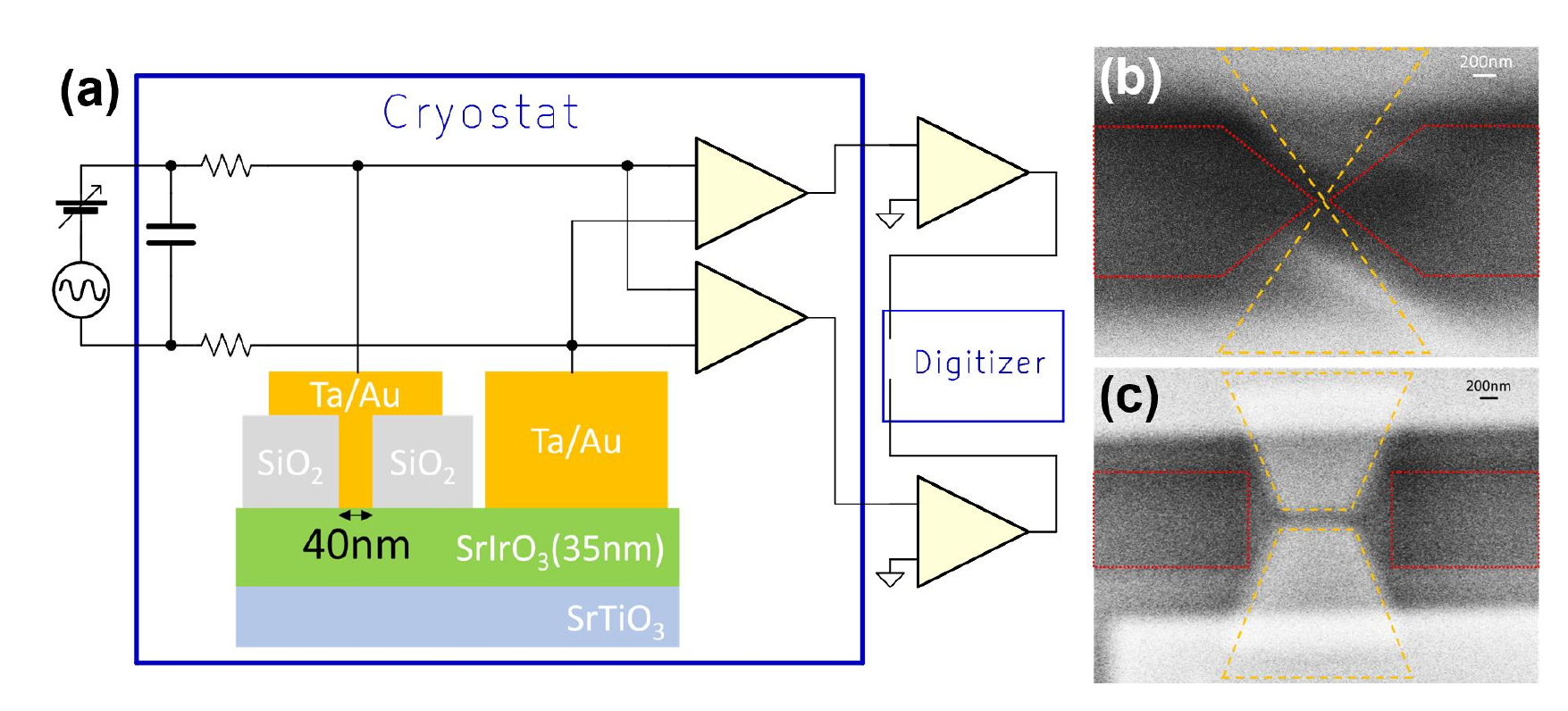}
	\end{subfigure}
	\begin{subfigure}[b]{0.45\textwidth}
		\includegraphics[width=\textwidth]{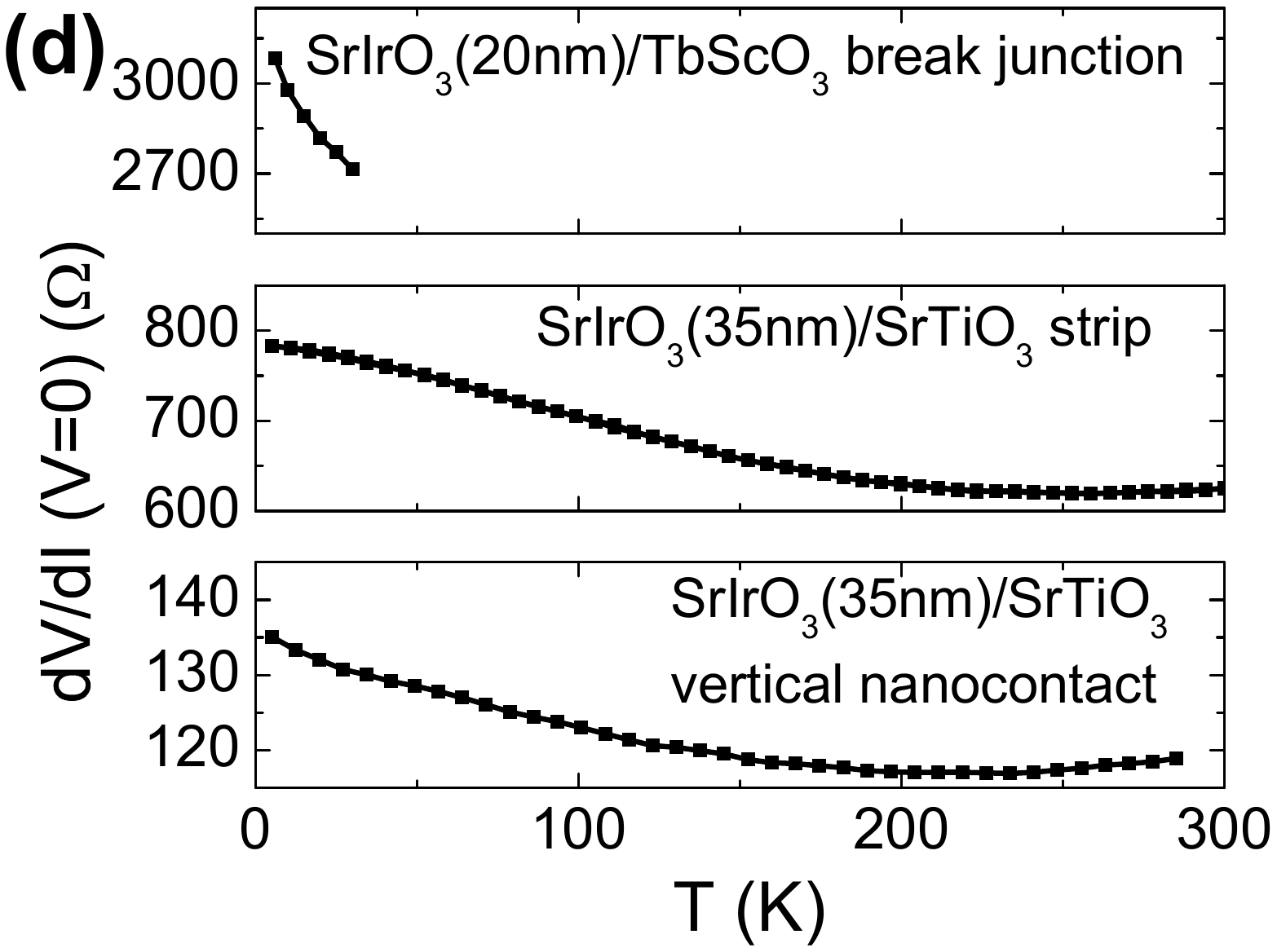}
	\end{subfigure}
	\begin{subfigure}[b]{0.45\textwidth}
		\includegraphics[width=\textwidth]{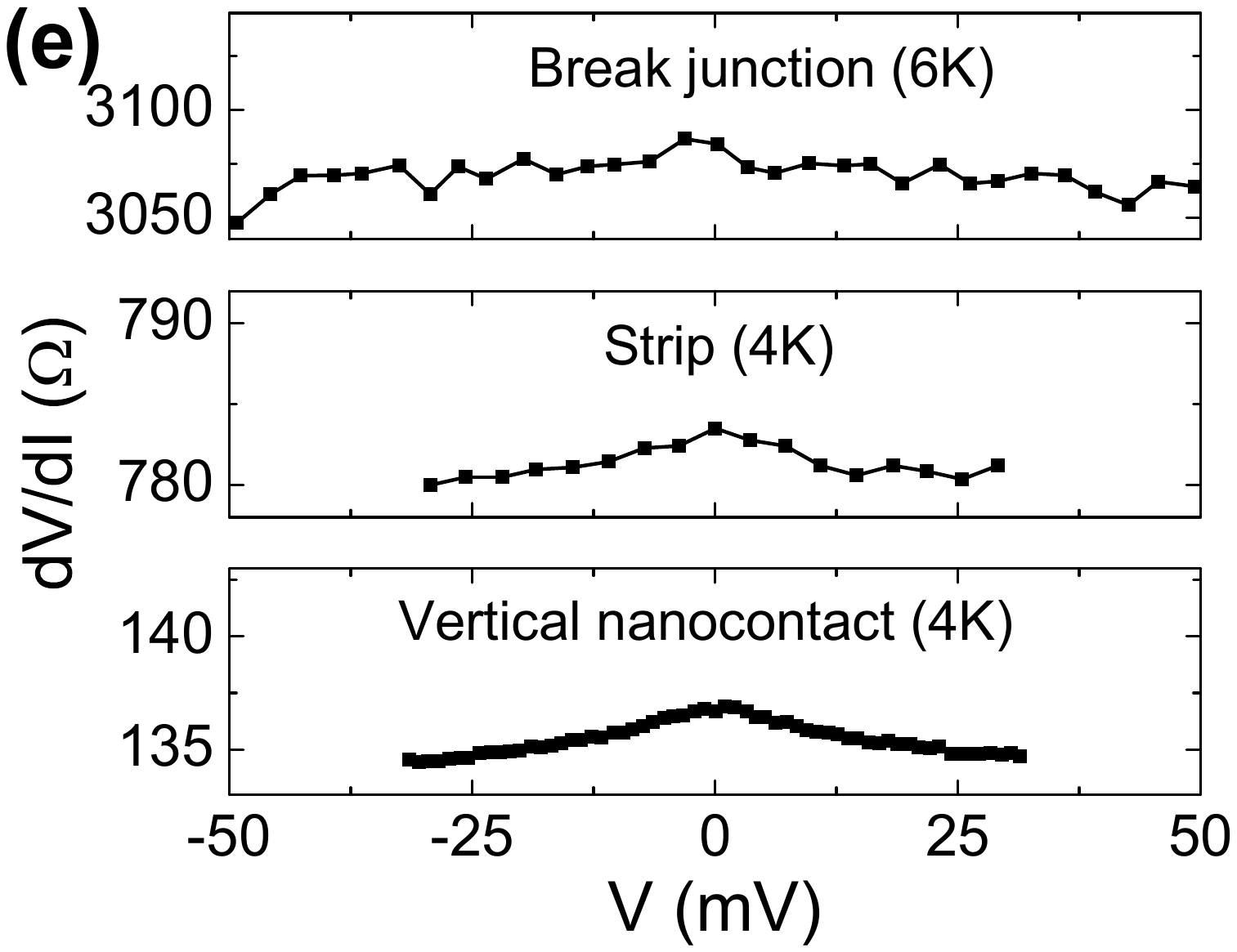}
	\end{subfigure}
	\caption{(a) Schematic of the vertical nanocontact and measurement circuit. Variable DC and AC ($100 \mu V$ at 100 Hz) voltage for resistance and SN measurements was sourced through two $12 k \Omega$ resistors. Noise was measured via two stages of parallel amplification (cryogenic first stage) followed by signal cross-correlation. (b), (c) SEM images of a break junction (b) and a $100$~nm long, $400$~nm wide strip (c). Yellow lines outline electrodes, red $-$ regions of SIO removed by Ar ion milling. (d), (e) Device resistance vs temperature (d) and bias voltage (e).}
	\label{fgr:fig2}
\end{figure}

The films were patterned into nanoscale metallic junctions using multistep e-beam lithography (see  Supplementary Materials for details). Junctions with length $L=100-500$~nm were lithographically-defined strips, Fig.\ref{fgr:fig2}(c). Shorter junctions were fabricated using two different  approaches. In a vertical nanocontact (VNC) with the size $40$~nm (Fig.\ref{fgr:fig2}(a)), current confinement defined junction length close to its diameter. In an alternative approach, break junctions with length $L\approx50$~nm were formed by electromigration starting with two connected sharp-point electrodes (Fig.\ref{fgr:fig2}(b)). SN in VNCs may be underestimated due to the resistance of the bottom SIO lead, while break junctions are prone to a large geometric uncertainty. Nevertheless,  consistency of results for two different junction geometries supports their validity.

The resistance $R(T)$ of the junctions was consistent with $\rho(T)$ for unpatterned films, showing that electronic properties of SIO were not compromised by nanopatterning, and the contribution of contact resistance was insignificant. For break junctions, bias was applied only at cryogenic $T<30$~K to avoid damage. Ohmic properties of devices were tested by the dependence of $R$ on bias. Resistance was almost constant for the break junctions, and slightly decreased with bias for other structures, Fig.\ref{fgr:fig2}(e). It is unlikely that the decrease is caused by Schottky contact barrier, which would have been larger in break junctions based on SIO(20) film since it is closer to the insulating state, and due to the small effective contact area. Regardless of the origin, the variation of $R$ is too small for the underlying transport process to significantly contribute to SN.

\begin{figure}
	\centering
	\begin{subfigure}[b]{0.45\textwidth}
		\includegraphics[width=\textwidth]{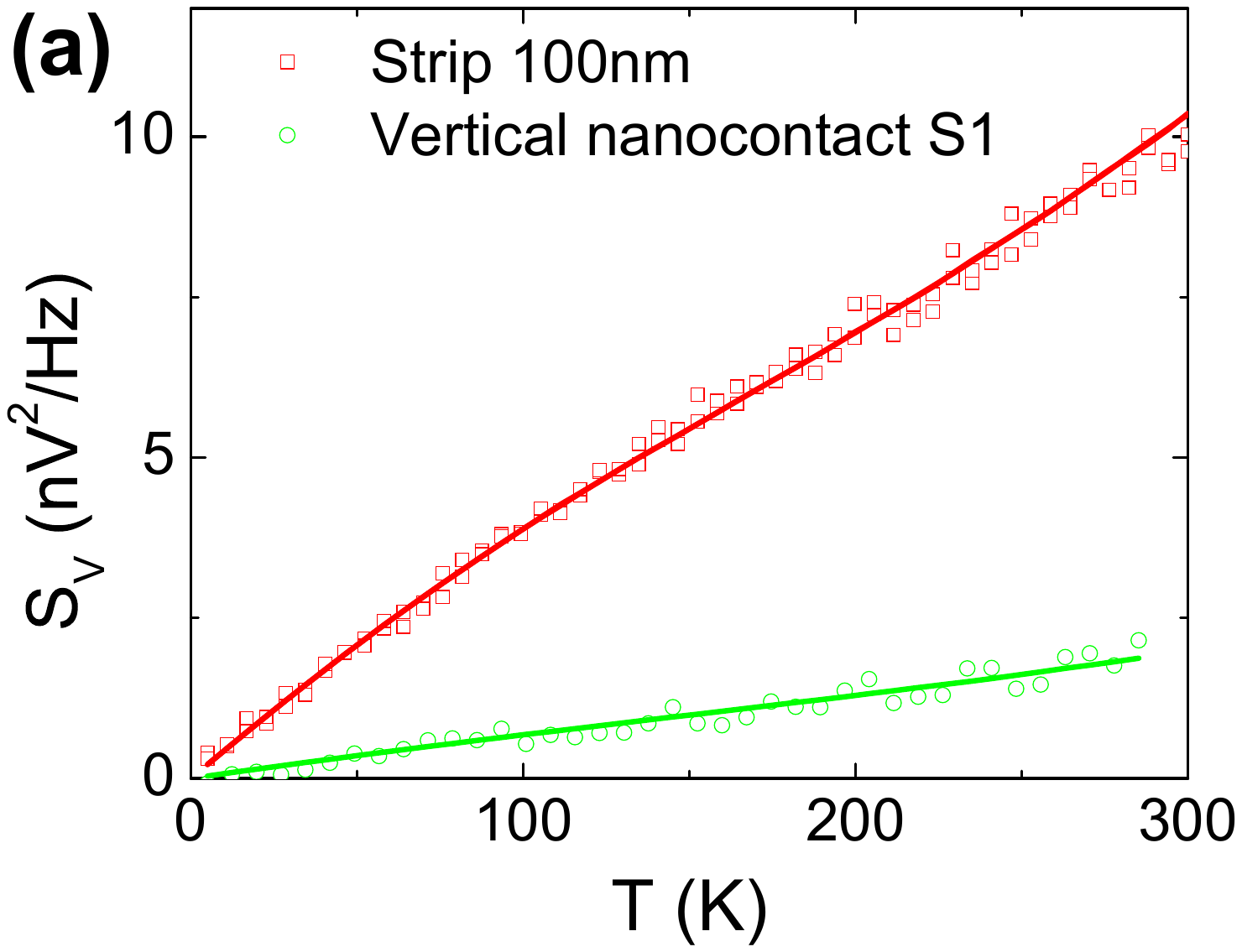}
	\end{subfigure}
	\begin{subfigure}[b]{0.45\textwidth}
		\includegraphics[width=\textwidth]{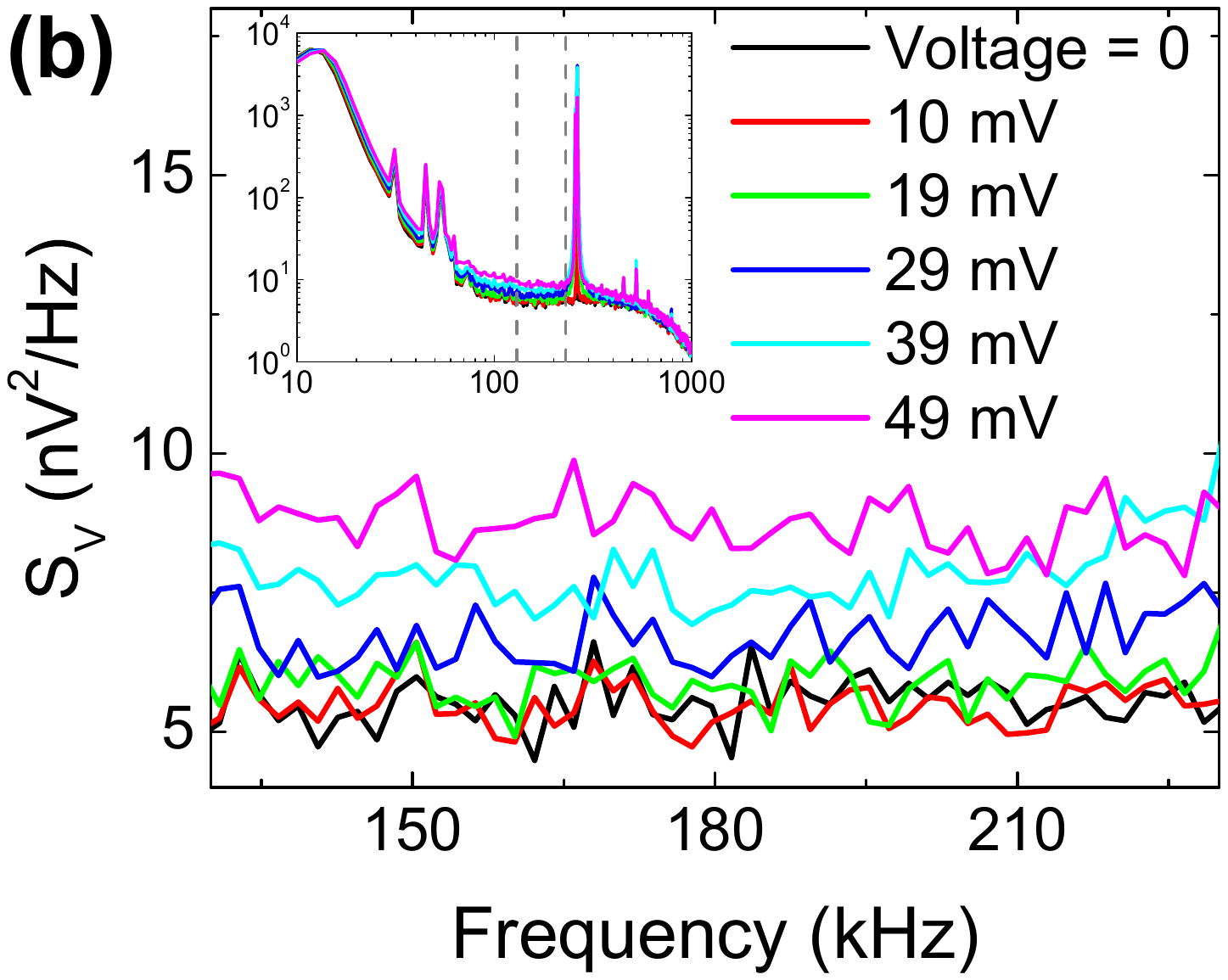}
	\end{subfigure}
	\begin{subfigure}[b]{0.45\textwidth}
		\includegraphics[width=\textwidth]{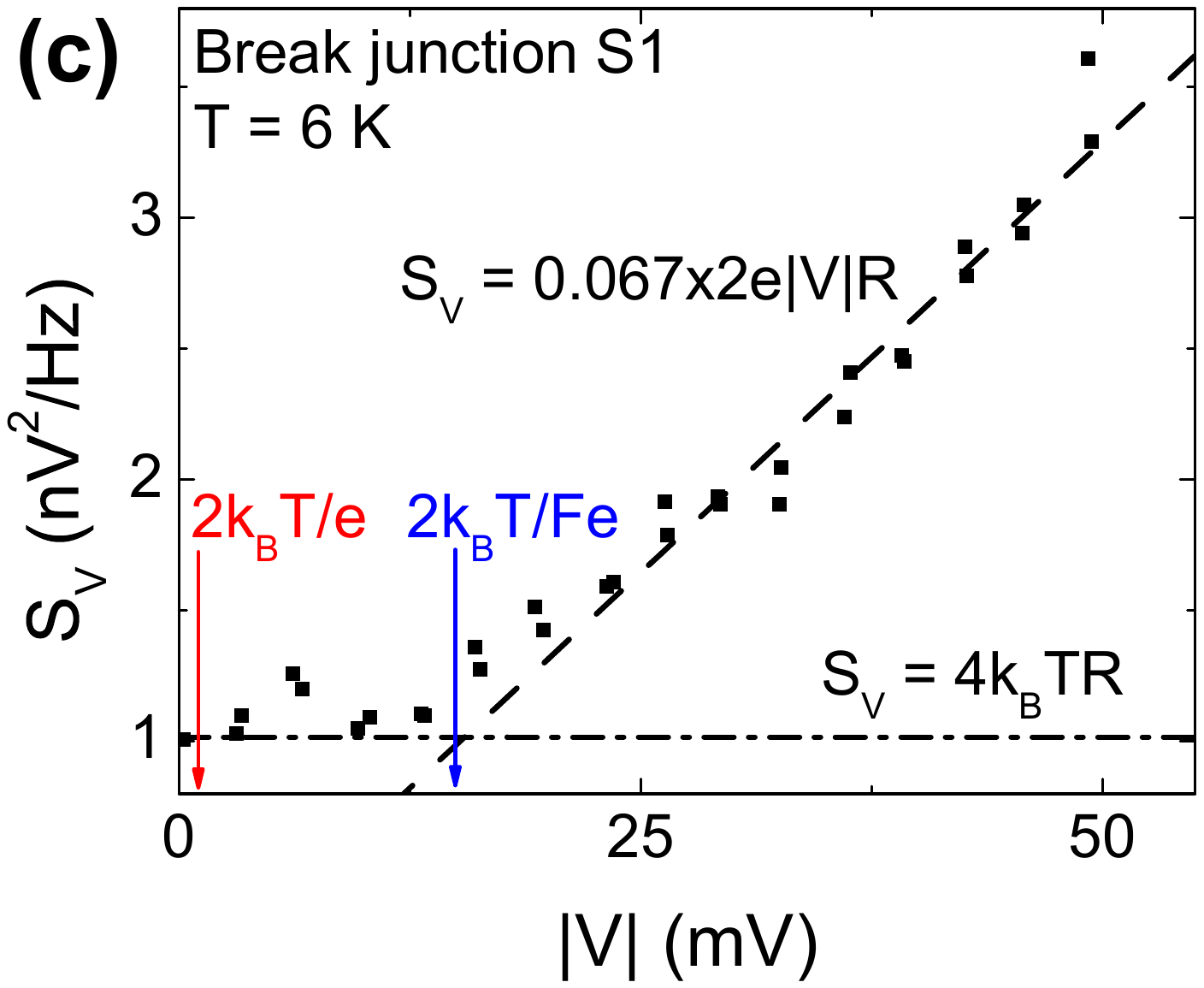}
	\end{subfigure}
	\begin{subfigure}[b]{0.45\textwidth}
		\includegraphics[width=\textwidth]{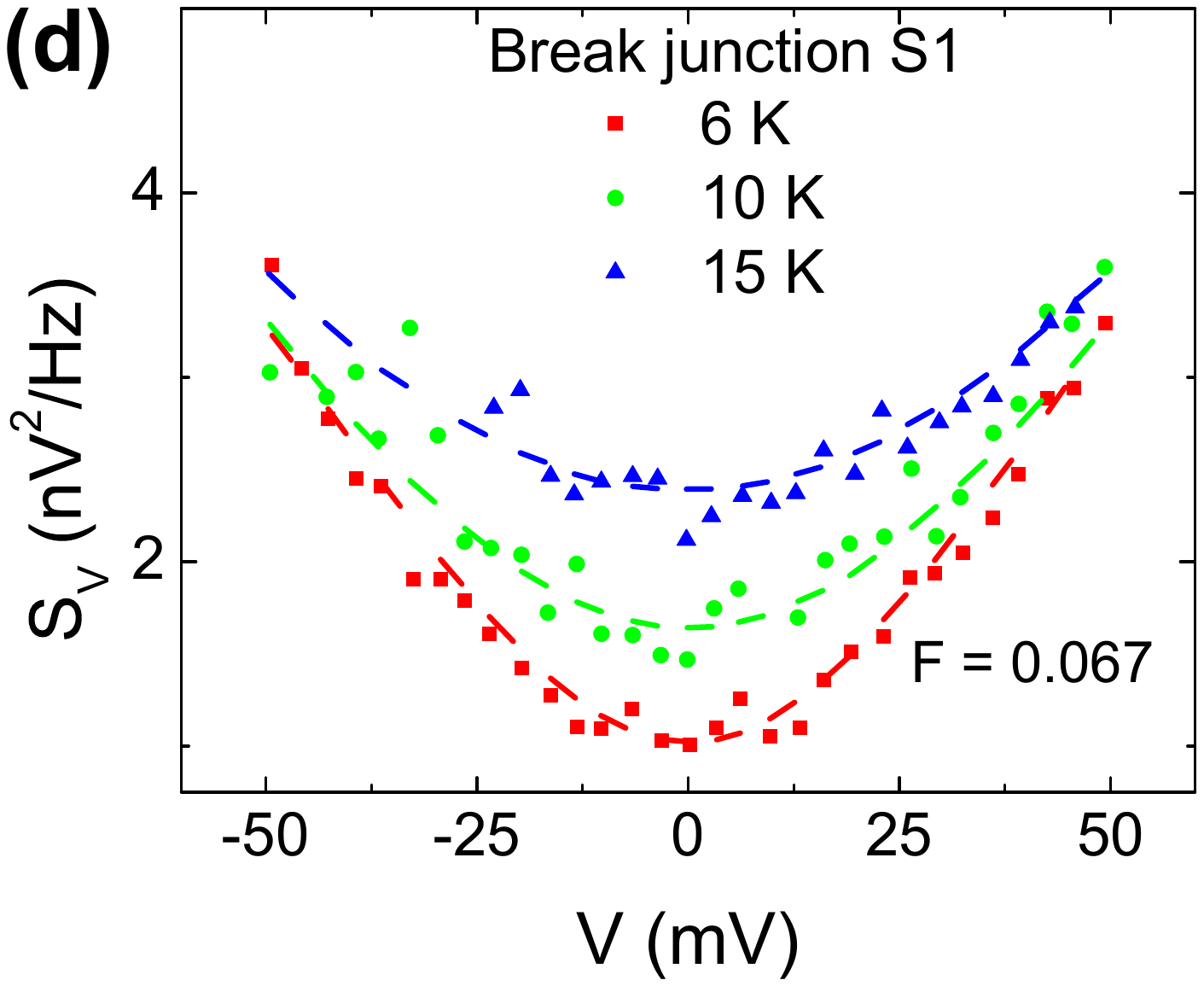}
	\end{subfigure}
	\begin{subfigure}[b]{0.45\textwidth}
		\includegraphics[width=\textwidth]{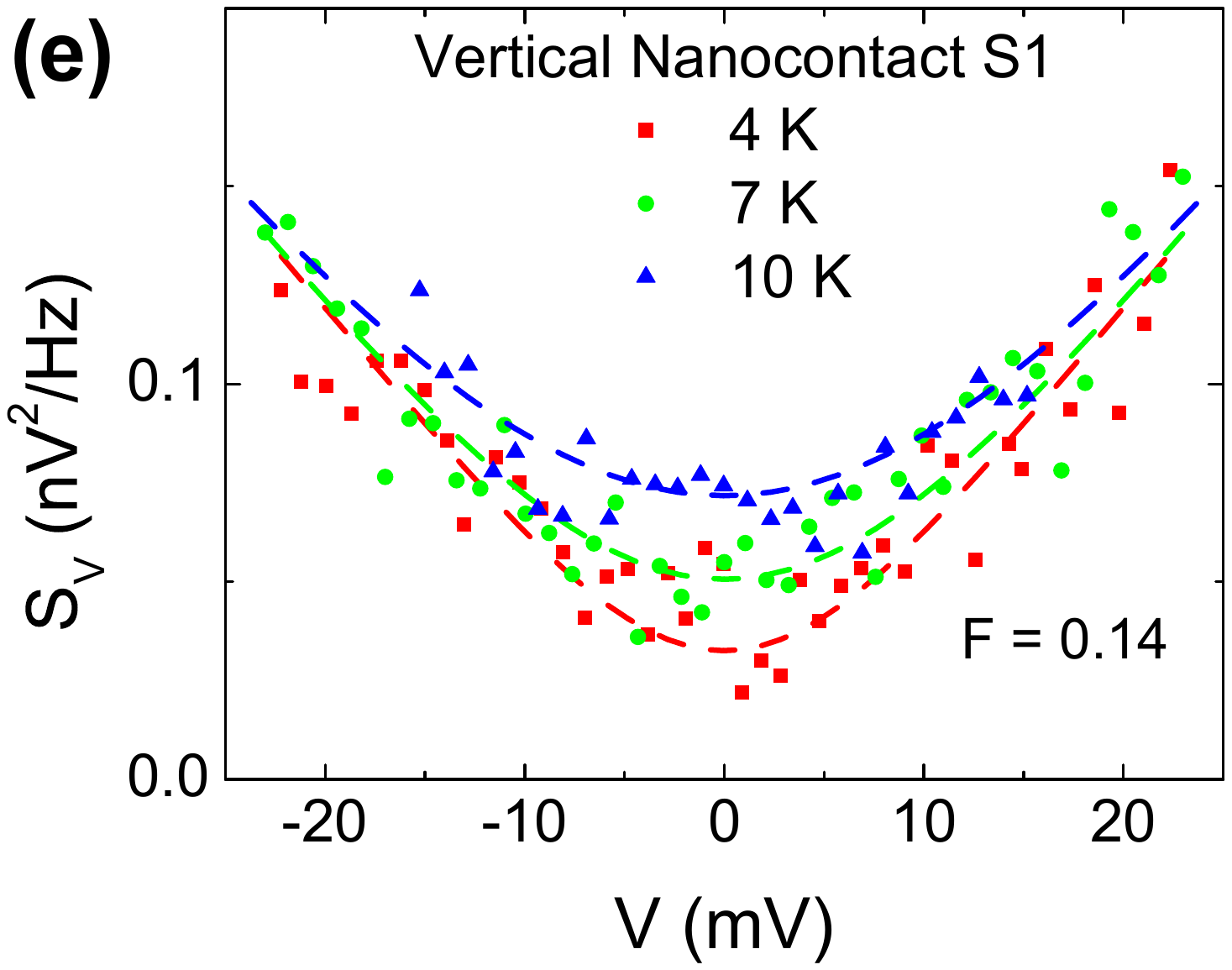}
	\end{subfigure}
	\begin{subfigure}[b]{0.45\textwidth}
		\includegraphics[width=\textwidth]{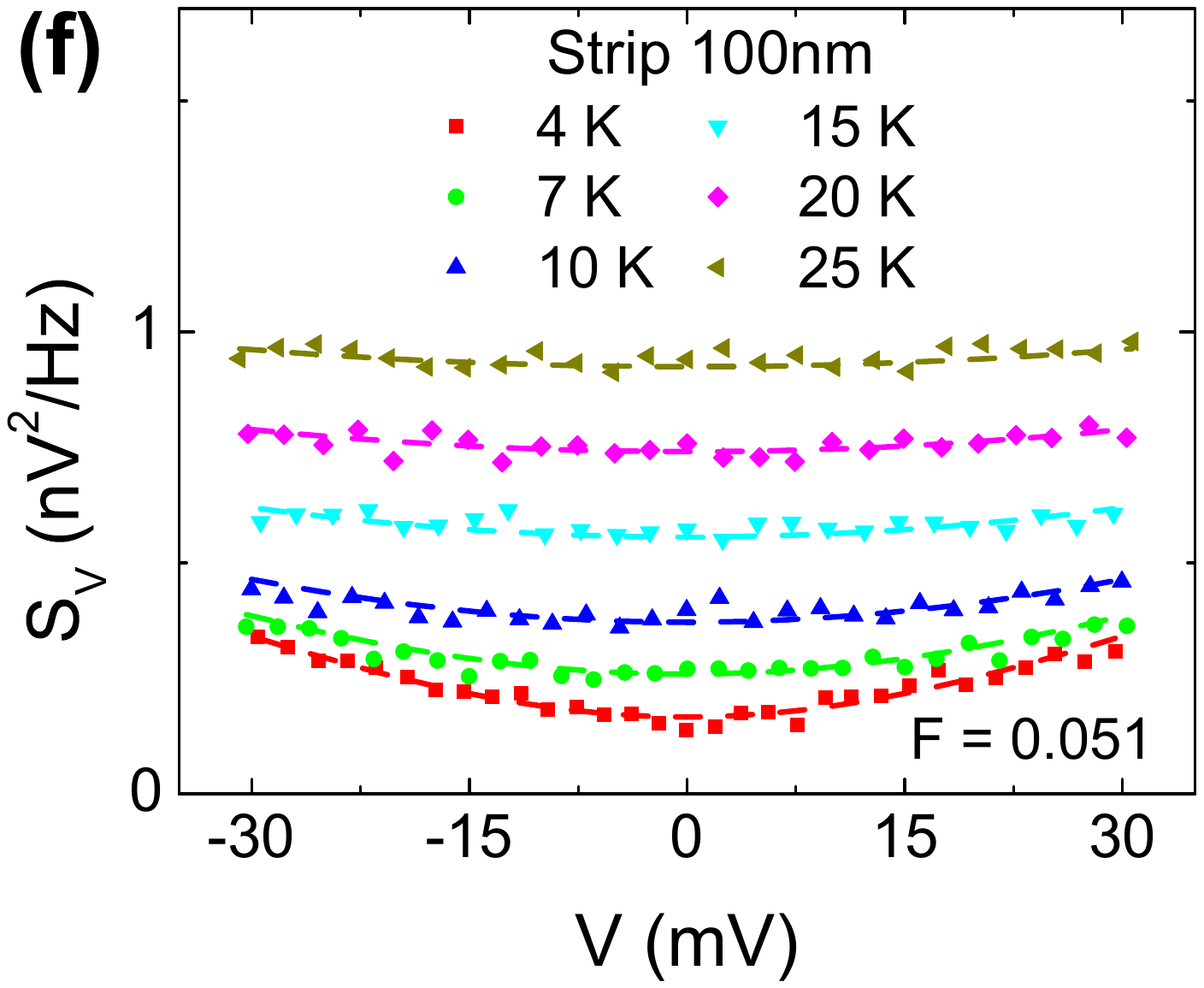}
	\end{subfigure}
	\caption{(a) Johnson noise vs $T$ for two SIO junctions. The solid lines are the expected values $S_V = 4 k_B T R(T)$. The nonlinearity is due to resistance variation with temperature, see Fig.\ref{fgr:fig2}(e). (b) Noise spectra of a break junction at different bias. Inset: broadband spectra showing EMI peaks and SN detection window. (c) $S_V$ vs $|V|$ for a break junction at $T=6$~K. The slope of linear dependence of $S_V$ on $V$ at large bias gives $F=0.067$. The crossover between Johnson noise and SN is $|V_{th}| = \frac{2 k_B T}{Fe}$. (d)-(f) $S_V$ vs $V$ for three different junctions. The dashed curves are fittings with Eq.~(\ref{eqn:shot-Johnson}).}
	\label{fgr:fig3}
\end{figure}

Our central result is the anomalous SN produced by the SIO nanojunctions under bias. Noise measurements were performed using the cross-correlation technique, Fig.\ref{fgr:fig2}(a) (see Supplementary Materials for details). 
The technique was validated by measurements of thermal (Johnson) noise vs $T$, which were in agreement with calculations without any adjustable parameters, Fig.\ref{fgr:fig3}(a). Electromagnetic interference (EMI) was manifested by sharp peaks in the noise spectra, which were the largest in the break junctions [inset in Fig.~\ref{fgr:fig3}(b)]. A frequency window where the detected noise was white and free from EMI was used in measurements.

Noise increased with bias while remaining white, as expected for SN and eliminating possible artifacts from  flicker noise, Fig.\ref{fgr:fig3}(b). The dependence of voltage noise $S_V$ on bias magnitude $|V|$ for one of the two studied break junctions is shown in Fig.~\ref{fgr:fig3}(c). Johnson noise dominates at small $V$, resulting in almost constant $S_V=4 k_B T  R$. SN is manifested by the linear dependence at large $V$, which can be described by 
\begin{equation}
  S_V = S_I  (\frac{\mathrm{d}V}{\mathrm{d}I})^2 =2Fe|V|R.
  \label{eqn:shot}
\end{equation}
The Fano factor $F=0.067$ is much smaller than $1/3$ expected for weak electron-electron interaction~\cite{nagaev1992shot,steinbach1996observation,kozub1995shot,nagaev1995influence} or $\sqrt{3}/4$ for electrons thermalized by interactions, consistent with SN suppression in a ``strange" metal~\cite{chen2023shot}. The role of thermal phonons is negligible at $T$ far below the Debye temperature, as confirmed by the lack of temperature dependence discussed below. Electron-phonon interaction would result in a nonlinear dependence $S_V \propto |V|^{\frac{2}{5}}$~\cite{nagaev1992shot,nagaev1995influence}, which is not observed. Therefore, in the single-particle Fermi liquid picture, suppressed SN cannot be explained by either electron-phonon or electron-electron interaction. 

Analysis of thermal effects provides insight into the mechanism of SN suppression. The crossover bias $V_{th}$ from Johnson noise to SN is defined by the intersection between the extrapolation of the linear dependence for SN and bias-independent Johnson noise, Fig.~\ref{fgr:fig3}(c). In the Landauer approximation, $V_{th}=\frac{2 k_B T}{e}$ is expected from the analysis of the Fermi-Dirac distribution in electrodes~\cite{beenakker1992suppression}. The same result is obtained in the diffusive approximation~\cite{nagaev1992shot,nagaev1995influence} from the analysis of thermal distribution in the junction. Instead, the observed thermal broadening is close to $\frac{2k_B T}{Fe}$, $15$ times larger than expected from these theories. 

An alternative model for electron transport is hopping through localized states in materials close to the insulating state~\cite{korotkov2000shot,sverdlov2001shot,danneau2010shot,kuznetsov2000partially,camino2003hopping,kinkhabwala2006numerical}, or their sequential tunneling through defective barriers~\cite{jiang2004low,zhang2019spin}. If the transmission process comprises $N$ hopping segments or tunneling events, the average bias voltage per segment is $V/N$, resulting in the Fano factor $F=1/N$ and thermal broadening $V_{th}=\frac{2N k_B T}{e}=\frac{2k_B T}{Fe}$ consistent with our data. By comparing fits obtained with several functional forms (see Supplementary Materials for details), we found that the entire dependence of noise on bias was best described by
\begin{equation}
  S_V = 2FeV R \coth (\frac{FeV}{2 k_B T}),
  \label{eqn:shot-Johnson}
\end{equation}
consistent with hopping transport.

Figures~\ref{fgr:fig3}(d)-(f) show $S_V$ vs bias for all three studied junction types. Despite different junction geometries, lengths, and values of $F$, Eq.~(\ref{eqn:shot-Johnson}) provided a good fit for all the junctions, with the effect of thermal broadening well-captured for all measurement temperatures (see supplementary Materials for additional data and fitting).
 
\begin{figure}
	\centering
	\begin{subfigure}[b]{0.45\textwidth}
		\centering
		\includegraphics[width=\textwidth]{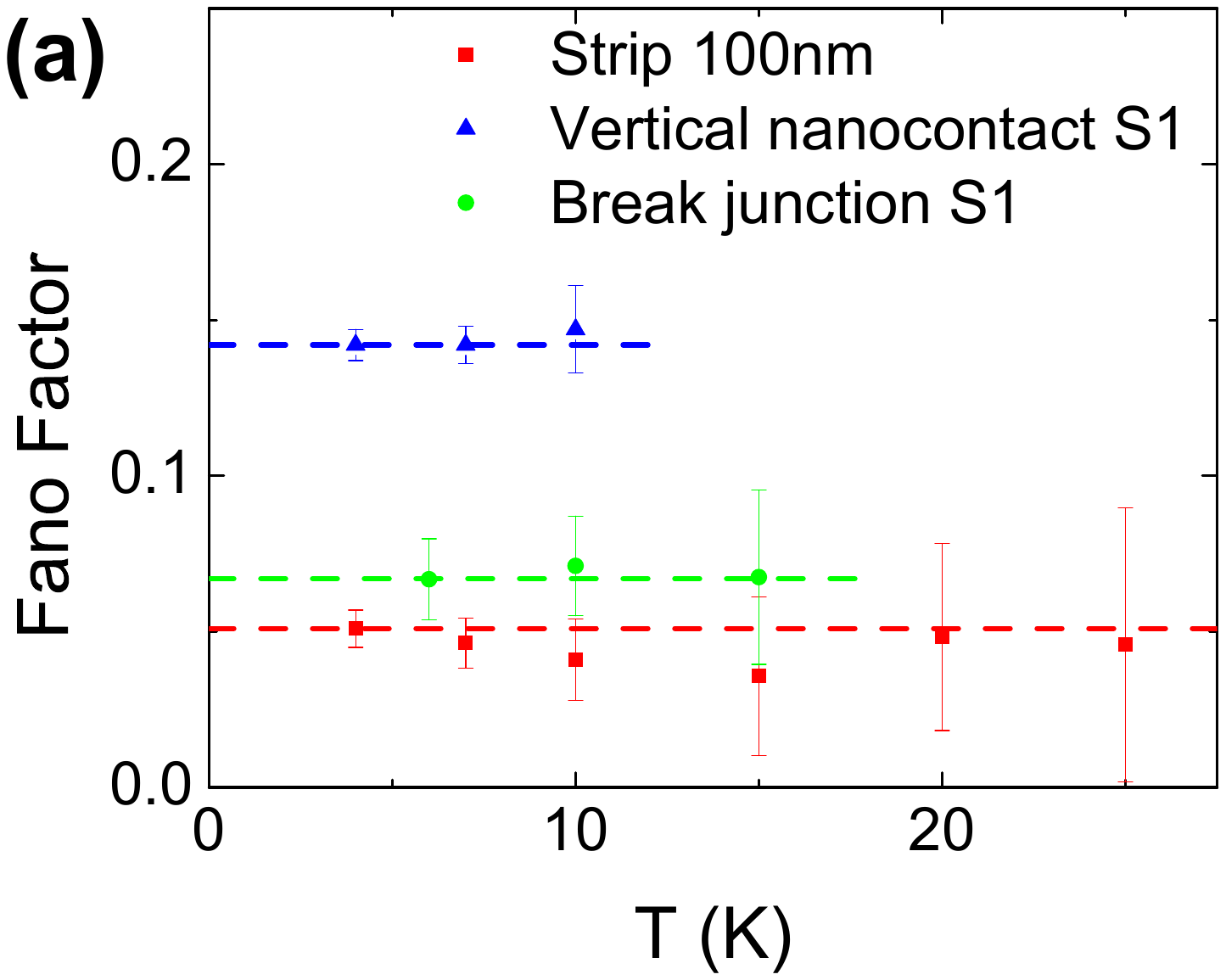}
	\end{subfigure}
	\begin{subfigure}[b]{0.45\textwidth}
		\centering
		\includegraphics[width=\textwidth]{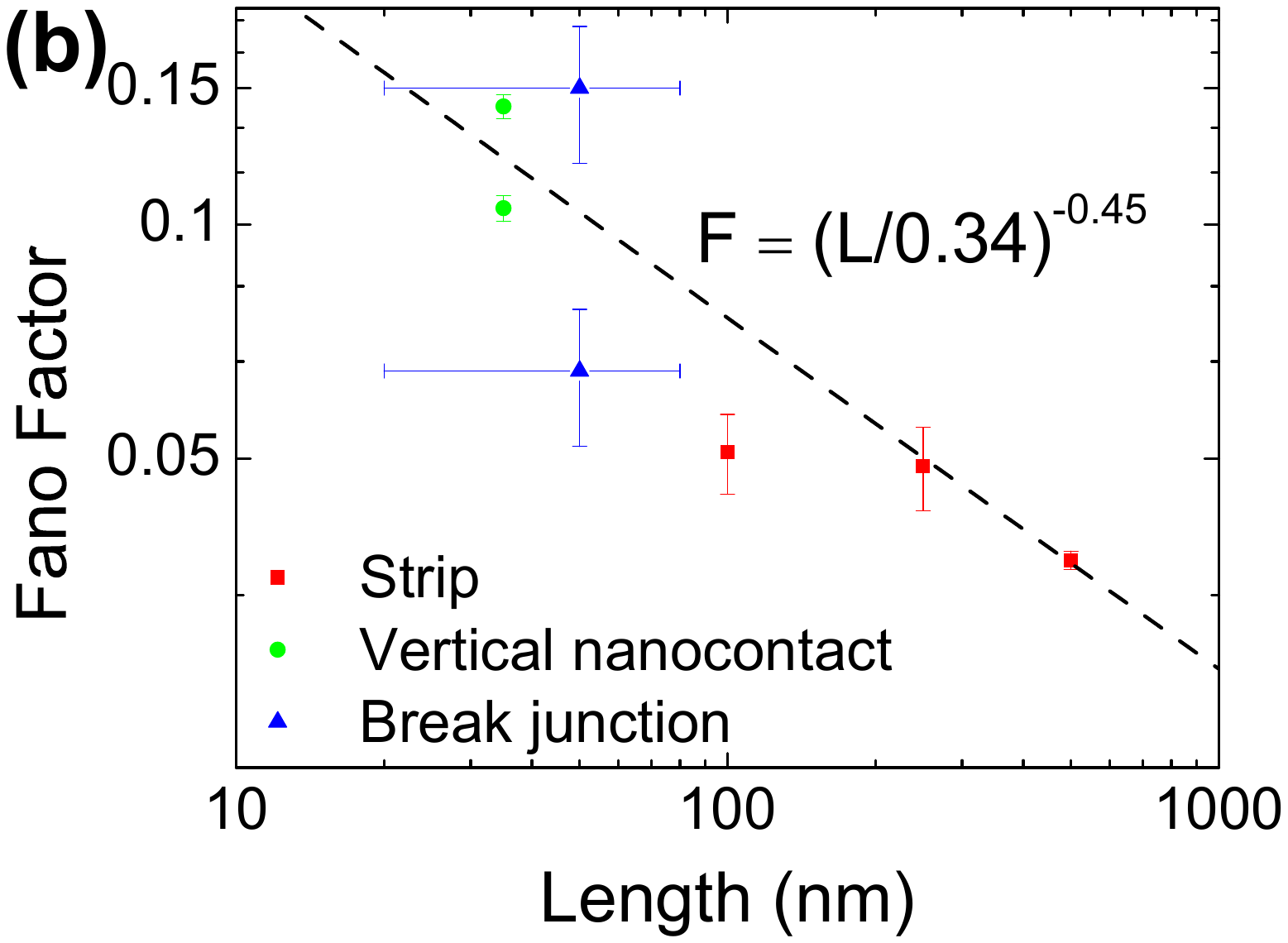}
	\end{subfigure}
    \begin{subfigure}[b]{0.9\textwidth}
		\centering
		\includegraphics[width=\textwidth]{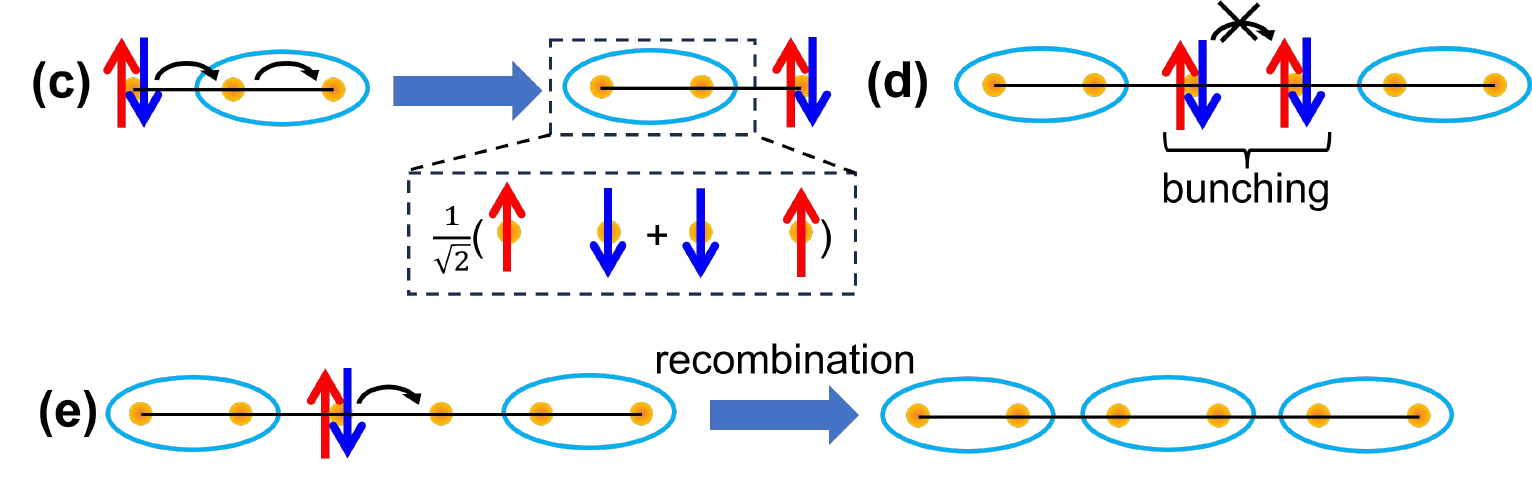}
	\end{subfigure}
	\caption{(a) Fano factors vs temperature for three junctions with different geometries. The dashed horizontal lines are guides to the eye. (b) Fano factors of all the measured junctions as a function of the junction length. Uncertainty of $F$ is estimated from numerical fitting, and uncertainty of the junction length $-$ from the SEM images and resistance. The dashed line shows the linear fitting of $\log(F)$ versus $\log(L)$, which gives $F = (\frac{L}{L_C})^{-\beta}$ with $L_c=0.34\pm0.05$~nm and $\beta=0.45\pm0.01$. (c)-(e) Schematics of quasi-localized charge hopping in the VB state. (c) Hopping of an excess electron (two electrons on site) through VB. (d) Hopping of an excess electron is blocked by another excess electron due to Pauli exclusion principle, causing bunching of hopping events. (e) Recombination of an excess electron and a hole results in hopping blockage for other charges. }
	\label{fgr:fig4}
\end{figure}
  
Figure~\ref{fgr:fig4}(a) shows the temperature dependence of the Fano factors extracted from fitting the noise data with Eq.~(\ref{eqn:shot-Johnson}). For clarity, only the three junctions shown in Fig.\ref{fgr:fig3}(d)-(f) are included. In the measured temperature range, the value of $F$ is constant for all the junctions within fitting uncertainty, confirming that scattering on phonons is not the origin of SN suppression.

The mechanism of hopping is elucidated by the dependence of the Fano factor on the junction length $L$, as shown for all the studied junctions in Fig.\ref{fgr:fig4}(b). The value of $F$ decreases with $L$, as expected due to  increasing number of hopping events. Linear fitting of $\log F$ versus $\log L$ gives $F = (\frac{L_c}{L})^{\beta}$, with the characteristic hopping length $L_c = 0.34\pm0.05$~nm and exponent $\beta = 0.45\pm0.01$. These values were extracted from just seven data points. Nevertheless, the agreement between the two contributions of $F$ to noise in Eq.~(\ref{eqn:shot-Johnson}) $-$ thermal broadening and slope $-$ gives us confidence that the determined values are not significantly affected by artifacts such as contact resistance.

The scaling with length is slower than $F\propto 1/L$ expected from independent charge hopping, while the hopping length close to the pseudo-cubic lattice constant $a=0.39$~nm of SIO~\cite{zhang2018review} is unphysically small for single-particle hopping in defect-dominated energy landscape. These anomalous behaviors can be explained by strong correlations, as discussed below.

\textbf{Analysis and discussion}. The lack of low-temperature resistivity divergence inconsistent with VRH~\cite{PhysRevB.44.3599} is commonly interpreted as evidence for diffusive charge transport in conducting SIO films.  However, our SN measurements demonstrate that conduction in the studied SIO films is mediated by hopping. We argue that single-particle VRH models are not applicable to SIO since its electronic properties are not described by the Fermi liquid picture~\cite{cao2007non,biswas2016metal,Sen2020}.  We propose a qualitative mechanism for the correlated transport in SIO based on the Hubbard model of a half-filled band with the effects of hopping $t$ somewhat dominant over the electron interaction $U$. We leave quantitative modeling and analysis of multi-orbital effects to future work.

If hopping is somewhat dominant over electron interaction, the insulating Mott state is suppressed, which may be facilitated by large SOC even if the Mott criterion for MIT is satisfied~\cite{zeb2012interplay}. Under these conditions, Fermi liquid may not be the ground state because of the energy reduction provided by the local Mott singlet correlations, which can be described as incoherent Cooper pairs~\cite{PhysRevB.91.115111}. 
In the resonating valence bond (RVB) model of the non-Fermi liquid state in doped Mott insulators, such pairs are approximated by valence bonds (VBs) $-$ singlets formed by electrons quasi-localized on neighboring sites.~\cite{doi:10.1126/science.235.4793.1196} In SIO, a relatively large ratio $t/U$ allows significant local charge fluctuations. Since VBs constrain site occupancy to a single electron, a quasi-localized excess electron or a hole can be located only on sites that do not participate in VBs. Consequently, charge diffusion is prevented by the dense VBs that limit the possible charge positions.

To develop a qualitative picture of charge transport in this state, consider a minimal three-site model of an excess electron next to a VB, Fig.~\ref{fgr:fig4}(c). The electron can move by exchanging places with the VB via double hopping. The final state has the same energy as the initial state, since they differ only by the relative positions of electron and VB. On the other hand, in the intermediate state with the extra electron on the middle site, nearest-neighbor singlet correlation is suppressed, resulting in a higher energy. Thus, this transport process is elastic electron tunneling governed by correlations, which in contrast to single-particle hopping or inelastic tunneling in disordered energy landscape does not lead to resistivity divergence at low $T$.

The characteristic hopping length $L_c = 0.34$~nm determined from the scaling of the Fano factor with $L$ is close to the pseudo-cubic lattice constant of $0.39$~nm~\cite{zhang2018review}. This value would be unphysically short in the single-electron picture, but is consistent with the proposed mechanism. The anomalously small scaling exponent can also be explained by correlations. Suppressed scaling exponent caused by electron bunching due to Pauli exclusion principle was predicted in simulations of hopping in confined geometries~\cite{korotkov2000shot,sverdlov2001shot,kinkhabwala2006numerical}. Similar bunching was observed in tunnel junctions as charge and spin blockage~\cite{zhang2019spin,garzon2007enhanced,bulka2000current,braun2006frequency}. In 2D systems, the effects of bunching are modest due to the existence of multiple hopping paths, resulting in the scaling exponents close to $\beta=1$~\cite{kinkhabwala2006numerical,kuznetsov2000partially}. On the other hand, strong bunching effects in 1D chains~\cite{korotkov2000shot} and quasi-1D systems~\cite{sverdlov2001shot} are predicted to yield $\beta=0.5$ remarkably close to $b=0.45$ observed in our measurements.

In our model of correlations in SIO, charge hopping is expected to be confined to quasi-1D paths by the dense arrangement of VBs, which should lead to efficient hopping blockage due to excess charge bunching, Fig.~\ref{fgr:fig4}(d). Furthermore, overlap of electron and hole hopping paths can result in their annihilation (Fig.\ref{fgr:fig4}(e)), creating a VB which reduces possible configurations of VBs that permit charge hopping. The latter mechanism may explain why the observed value $\beta=0.45$ is even smaller than $0.5$ expected for 1D hopping of a single type of charge carriers. The scaling exponent may be also reduced by the strong correlations beyond the effective single-particle hopping picture.

The proposed model can explain the thickness dependence of transport properties in moderately thin films. Increased geometric confinement at smaller film thickness leads to larger overlap between charge hopping paths, increasing the probability of blocking via the mechanisms discussed above. In contrast to single charge swap with a VB, these processes involve interactions between excess charges and do not necessarily conserve energy, resulting in an increase of resistivity at low $T$~\cite{fuentes2019resistive,zhang2018review,zhang2020abrupt}. A similar mechanism can explain the evolution of electronic properties across the Sr$_{n+1}$Ir$_{n}$O$_{3n+1}$ iridate family with $n>2$, before the onset of Mott state.

In summary, we studied shot noise in metallic SrIrO$_3$ nanojunctions with lengths between $40$~nm and $500$~nm. All the studied junctions exhibit strong shot noise suppression which cannot be explained by electron-electron or electron-phonon interaction in the single-particle Fermi liquid picture. Analysis of thermal broadening of noise dependence on bias shows that conduction occurs not via the commonly assumed electron diffusion, but rather by charge hopping. The dependence of shot noise on the junction length reveals the dominant role of electron correlations. We propose that the unusual electronic properties of SrIrO$_3$ films can be explained by charge hopping along quasi-1D conduction paths confined by Mott singlet correlations. This mechanism lifts the constraints imposed by energy conservation on single-particle hopping, avoiding resistivity divergence at low temperatures. The proposed mechanism may be relevant to one of the puzzling hallmark features of the ``strange" metal state observed in other materials close to Mott transition such as high-temperature superconductors $-$ the non-saturating linear temperature dependence of resistivity~\cite{PhysRevLett.127.086601}. Our study elucidates the nature of the non-Fermi liquid state in SrIrO$_3$, and advances shot noise measurements as a powerful technique for the characterization of quantum materials. 

\begin{acknowledgement}
Y.Z and S.I acknowledge support by the NSF award ECCS-2005786, the Tarbutton fellowship, and the SEED award from the Research Corporation for Science Advancement. J.L. acknowledges support from the National Science Foundation under Grant No. DMR-1848269. S.P. acknowledges funding from the State of Tennessee and Tennessee Higher Education Commission through the Center for Materials Processing.
\end{acknowledgement}

\begin{suppinfo}

\end{suppinfo}

\bibliography{SrIrO3}

\end{document}